\documentclass[12pt,preprint]{aastex}
\slugcomment{To appear in The Astrophysical Journal}

\shorttitle{Bardeen-Petterson effect in the inner disk of NGC\,1068}
\shortauthors{Caproni et al.}

\begin{document}


\title{Bardeen-Petterson effect and the disk structure of the Seyfert galaxy NGC\,1068}


\author{Anderson Caproni and Zulema Abraham}
\affil{Instituto de Astronomia, Geof\'\i sica e Ci\^encias
Atmosf\'ericas, Universidade de S\~ao Paulo, R. do
Mat\~ao 1226, Cidade Universit\'aria, CEP 05508-900, S\~ao Paulo, SP,
Brazil; acaproni@astro.iag.usp.br, zulema@astro.iag.usp.br.}

\and

\author{Herman J. Mosquera Cuesta \altaffilmark{1}}
\affil{Instituto de Cosmologia, Relatividade e Astrof\'{\i}sica 
(ICRA-BR),  Centro Brasileiro de Pesquisas F\'\i sicas, R. Dr. Xavier 
Sigaud 150, CEP 22290-180, Rio de Janeiro, RJ, Brazil; hermanjc@cbpf.br.}

%%\email{aastex-help@aas.org}

%%\and

%%\author{R. J. Hanisch\altaffilmark{5}}
%%\affil{Space Telescope Science Institute, Baltimore, MD 21218}


\altaffiltext{1}{Abdus Salam International Centre for Theoretical Physics, 
Strada Costiera 11, Miramare 34014, Trieste, Italy}

%%\altaffiltext{2}{Society of Fellows, Harvard University.}
%%\altaffiltext{3}{present address: Center for Astrophysics,
%%    60 Garden Street, Cambridge, MA 02138}
%%\altaffiltext{4}{Visiting Programmer, Space Telescope Science Institute}
%%\altaffiltext{5}{Patron, Alonso's Bar and Grill}


\begin{abstract}
VLBA high spatial resolution observations of the disk structure of the
active galactic nucleus NGC 1068 has recently revealed that the
kinematics and geometry of this AGN is well characterized by an outer
disk of H$_2$O maser emission having a compact milliarcsecond (parsec)
scale structure, which is encircling a thin rotating inner disk
surrounding a $\sim$10$^7$ M$_\sun$ compact mass, likely a black
hole. A curious feature in this source
is the occurrence of a misalignment between the inner and outer parts 
of the disk, with the galaxy's radio jet being orthogonal to the
inner disk. We interpret this peculiar configuration as due to the
Bardeen-Petterson effect, a general relativistic effect that warps
an initially inclined (to the black hole equator) viscous disk, and
drives the angular momentum vector of its inner part into alignment
with the rotating black hole spin. We estimate the time-scale for both 
angular momenta to get aligned as a function the spin parameter of the 
Kerr black hole. We also  reproduce the shape of the parsec and kiloparsec 
scale jets, assuming a model in which the jet is precessing with a 
period and aperture angle that decrease exponentially with time, as expected 
from the Bardeen-Petterson effect.
\end{abstract}


%% subject header.  Objects should be in the appropriate "individual"
%% headers (e.g. quasars: individual, stars: individual, etc.) with the
%% additional provision that the total number of headers, including each
%% individual object, not exceed six.  The \objectname{} macro, and its
%% alias \object{}, is used to mark each object.  The macro takes the object
%% name as its primary argument.  This name will appear in the paper
%% and serve as the link's anchor in the electronic edition if the name
%% is recognized by the data centers.  The macro also takes an optional
%% argument in parentheses in cases where the data center identification
%% differs from what is to be printed in the paper.



\section{Introduction}

Recent mid-infrared interferometric observations \citep{jaf04} were able 
to resolve the inner dust torus of the nearby (14.4 Mpc; \citealt{bla97}) 
Seyfert 2 galaxy NGC\,1068. The emission distribution along this torus 
showed a small hot source (probably the accretion disk), surrounded by 
a parsec-scale warm dust structure. 

The hot source coincides with the bright H$_2$O masers 
(\citealt{gre96,gal96b,gal01,gal04}) detected with Very Long Baseline 
Interferometry (VLBI). The maser emission seems to originate in a 
sub-Keplerian rotating disk with inner and outer radius corresponding 
to $\sim$0.65 and 1.1 pc respectively enclosing, within its inner radius, 
a black hole and an accretion disk of  $\sim$ 10$^7$ M$_\sun$. The 
departure from Keplerian motion has been attributed to the high mass 
of the accretion disk that dilutes the gravitational field of the 
black hole (e.g., \citealt{hure02,lobe03}). 

The continuum emission at sub-arcsecond scales revealed four components 
arranged in a jet-counterjet structure, with the core coinciding with 
the accretion disk position (\citealt{wiul87,mux96,gal96a,gal04}). The 
radio jet orientation changes monotonically with its distance from the 
core, bending towards the northeast direction. At larger scales, extended 
lobes are observed, with the northeast lobe having a conic shape, 
similar to the bow shock structure produced by jet-environment 
interaction \citep{wiul87}.

An interesting result from maser observations is that the maser disk 
is not completely aligned with the major axis of the accretion disk 
across all its extension (e.g., \citealt{grgw97,gal04}). Indeed, 
they reveal a maser distribution similar to a warped disk, with 
the inner parts almost perpendicular to the jet, but deviating 
from this configuration as the distance to the core increases. 
\citet{gal04} proposed a configuration in which the thin hot 
disk is perpendicular to the radio jet, while the misaligned 
maser disk points towards the parsec-scale dusty-torus.

In this work we investigate the possibility that both the bent jet 
and warped accretion disk are a consequence of the Bardeen-Petterson 
effect, which results from the misalignment between the angular momenta 
of the Kerr black hole and disk. In $\S$ 2, we discuss the 
Bardeen-Petterson mechanism, used to calculate the time-scale for 
alignment between the two angular momenta. In $\S$ 3 we present the 
physical parameters of the accretion disk of NGC\,1068, as constrained 
by the observations found in the literature, and we use them to model 
the warping of the accretion disk due to the Bardeen-Petterson effect 
as a function of the alignment time-scale. In $\S$ 4, we show how 
jet morphology can be used independently to estimate the alignment 
time-scale. Conclusions are presented in $\S$ 5.

\section{The Bardeen-Petterson effect and the alignment time-scale}

Frame dragging produced by a Kerr black hole causes precession of a 
particle if its orbital plane is inclined in relation to the equatorial 
plane of the black hole. The precession angular velocity $\Omega_{\mathrm{LT}}$ 
due to the Lense-Thirring effect is given by (e.g., \citealt{wilk72}):

\begin{equation}
\Omega_\mathrm{LT} = \frac{2G}{c^2}\frac{J_\mathrm{BH}}{r^3},
\end{equation}
\\where $G$ is the gravitational constant, $c$ is the light speed, $r$ is the 
radial distance from a black hole of mass $M_\mathrm{BH}$ and angular  momentum 
$J_\mathrm{BH}$, defined as $J_\mathrm{BH}=a_\ast GM_\mathrm{BH}^2/c$, with 
$a_\ast$  the ratio between the actual angular momentum and its maximum possible 
value.

The combined action of the Lense-Thirring effect and the internal viscosity 
of the accretion disk  forces the alignment between the angular momenta of 
the Kerr black hole and the accretion disk. This is known as the Bardeen-Petterson 
effect \citep{bape75} and affects only the innermost part of the disk, because of  
the short range of the Lense-Thirring effect, while its outer parts tend to remain in its original configuration. The %%@
transition radius between 
these two regimes is known as Bardeen-Petterson radius $R_\mathrm{BP}$, shown 
schematically in Figure 1; its exact location depends mainly on the physical 
properties of the accretion disk (\citealt{bape75,kupr85,ivil97,nepa00,lub02,fran05}). 

\begin{figure}
%%\epsscale{.80}
\plotone{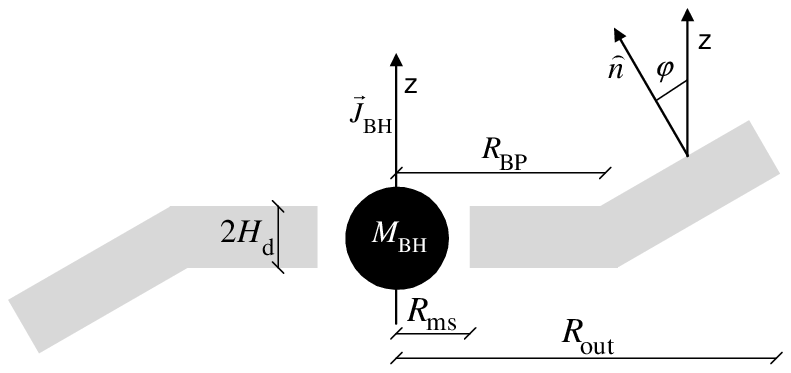}
\caption{Schematic view of the Bardeen-Petterson effect. An accretion 
disk with inner and outer radius $R_\mathrm{ms}$ and $R_\mathrm{out}$ respectively, 
having a semi-thickness $H_\mathrm{d}$ and misaligned initially by an angle 
$\varphi$ in relation to the angular momentum of the black hole $J_\mathrm{BH}$ 
and mass $M_\mathrm{BH}$, will be warped by the Bardeen-Petterson effect. The 
Bardeen-Petterson radius $R_\mathrm{BP}$ marks the transition between the aligned 
and misaligned disk in relation to the black hole's equator.}
\label{Warped disk}
\end{figure}

\placefigure{Warped disk}

The time-scale for alignment can be calculated as (e.g., \citealt{naar99}):

\begin{equation}
T_\mathrm{align} =
 J_\mathrm{BH}\left(\frac{dJ_\mathrm{BH}}{dt}\right)^{-1}\sin\varphi,
\end{equation}
\\where $\varphi$ is the  angle between the black hole spin axis and the  direction perpendicular  to  the outer 
disk.
The time derivative of $J_\mathrm{BH}$ has the form:

\begin{equation}
\frac{dJ_\mathrm{BH}}{dt} =
 -2\pi\sin\varphi\int_{R_\mathrm{BP}}^{R_\mathrm{out}}\Omega_\mathrm{LT}(r)L_\mathrm{d}(r)rdr,
\end{equation}
\\where $R_\mathrm{out}$ is the outer disk  radius  and 
$L_\mathrm{d}(r)=\Sigma_\mathrm{d}(r)\Omega_\mathrm{d}(r)r^2$ is its 
differential angular momentum (e.g., \citealt{cap04}). $\Omega_\mathrm{d}$ 
is the disk angular velocity and $\Sigma_\mathrm{d}$ is the surface 
density of the accretion disk integrated over its semi-thickness 
$H_\mathrm{d}$, derived following \citet{saco81} (see also \citealt{bar98}): 

\begin{equation}
H_\mathrm{d}(r) =
 \frac{H_\mathrm{nsg}(r)H_\mathrm{sg}(r)}{\sqrt{H_\mathrm{nsg}^2(r)+H_\mathrm{sg}^2(r)}}
\end{equation}
\\where $H_\mathrm{nsg}=c_\mathrm{s}/\Omega_\mathrm{d}$, 
$H_\mathrm{sg}=c_\mathrm{s}^2/(\pi G\Sigma_\mathrm{d})$ and
$c_\mathrm{s}$ is the sound speed, defined as:

\begin{equation}
c_\mathrm{s}(r) = \sqrt{-\Gamma\frac{d\ln\Omega_\mathrm{d}(r)}{d\ln r}
\frac{\nu_1(r)\Omega_\mathrm{d}(r)}{\alpha}},
\end{equation}
\\where $\Gamma$ is the politropic index of the  
gas, which we have assumed equal to 5/3.

A rough estimate of $R_\mathrm{BP}$ can be obtained comparing the 
time-scales for Lense-Thirring precession and warp transmission through 
the disk (e.g., \citealt{naar99}) that, on the other hand, will depend on 
how the warps are being communicated along it. If they are transmitted 
diffusively: 

\begin{equation}
R_\mathrm{BP}^\mathrm{diff}=
\sqrt{\nu_2/\Omega_\mathrm{LT}},
\end{equation} 
where $\nu_2$ is the viscosity acting on the  direction perpendicular to 
the disk; both parameters must be calculated at  $R_\mathrm{BP}^\mathrm{diff}$.
If the time-scales for the warp and surface density evolution are similar  $\nu_2\sim \nu_1$, where $\nu_1$ is the %%@
viscosity along the disk; otherwise $\nu_2\sim f(\alpha)\nu_1$, 
where $f(\alpha)$ is a function of the dimensionless viscosity parameter $\alpha$ 
introduced by \citet{shsu73};  we  adopted 
$f(\alpha)=2(1+7\alpha^2)/[\alpha(4+\alpha^2)]$ as derived by \citet{ogil99}. 
Assuming that the inner radius of the accretion disk is   the marginally stable 
orbit $R_\mathrm{ms}$,  the viscosity $\nu_1$ can be written as:

\begin{equation}
\nu_1 = -\frac{\dot{M}}{2\pi\Sigma_\mathrm{d}(r)}\left[\frac{d\ln\Omega_\mathrm{d}(r)}{d\ln
 r}\right]^{-1}\left[
 1-\left(\frac{R_\mathrm{ms}}{r}\right)^2\frac{\Omega_\mathrm{d}(R_\mathrm{ms})}{\Omega_\mathrm{d}(r)}
 \right],
\end{equation}
\\$\dot{M}$ is the accretion rate, which is related to the bolometric luminosity $L_{\rm bol}$ through 
$\dot{M}/\dot{M}_{\rm Edd}=L_{\rm bol}/L_{\rm Edd}$, where $\dot{M}_{\rm Edd}$ and $L_{\rm Edd}$ are the Eddington mass %%@
accretion rate and luminosity, respectively.

\begin{figure*}
%%\epsscale{.80}
\plotone{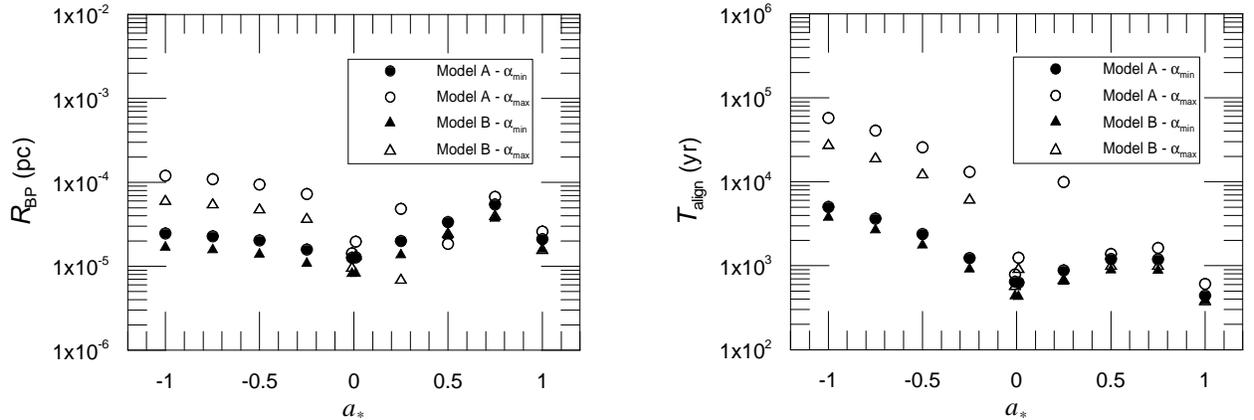}
\caption{{\it Left panel}: Bardeen-Petterson radius as a function of the black hole 
spin for NGC\,1068. Each point corresponds to the mean value of $R_\mathrm{BP}$ 
considering the limits of minimum and maximum accretion rate. Circles and triangles 
refer respectively to the models A and B. Filled symbols are related to the lower 
value of $\alpha$-parameter, while open ones to its upper value. {\it Right panel}: 
Time-scale for alignment between accretion disk and the black hole's equator, using 
the same nomenclature for the symbols.}
\label{RBPandTalign}
\end{figure*}

\placefigure{RBPandTalign}

In the wave-like regime:

\begin{equation}  
R_\mathrm{BP}^\mathrm{w} = c_\mathrm{s}/\Omega_\mathrm{LT}.
\end{equation}
 
The transition from the diffusive to wave-like regime occurs at a radius 
$R_\mathrm{T}\sim H_\mathrm{d}/\alpha$  \citep{pali95}.

\section{Application to the accretion disk of NGC\,1068}

The physical parameters of the black hole accretion disk system of NGC\,1068 were 
determined by \citet{hure02} and \citet{lobe03} from the analysis of the maser 
line velocities; they are presented in Table 1 and were used in our calculations 
of the Bardeen-Petterson radius and the alignment time-scale.  

The maser velocities present signatures of sub-Keplerian motion; however, in  equation (3) we used the relativistic %%@
Keplerian angular velocity 
\citep{abr78}, since at the radius in which this assumption is not valid, 
the angular momentum $L_{\rm d}$ becomes negligible. 

We assumed a power-law surface density distribution for the accretion disk 
$\Sigma(r)=\Sigma_0(r/R_\mathrm{g})^s$, where $R_\mathrm{g}=GM_\mathrm{BH}/c^2$ 
is the gravitational radius and  $s=-1.05$ \citep{hure02}. The constant $\Sigma_0$ 
was determined from the mass of the disk $M_\mathrm{d}$, derived by \citet{hure02} 
and \citet{lobe03}, integrating  $\Sigma(r)$ from the inner to the outer disk radius. 

The $\alpha$-parameter was obtained from the expression 
$\dot M =(28.1 \pm 0.2)\alpha$ M$_\odot$ yr$^{-1}$ given by \citet{lobe03}. 
The accretion rate can be calculated from the bolometric luminosity 
$L_{\rm bol}=\epsilon{\dot M}c^2$, given 
the accretion efficiency $\epsilon$, which depends on the black hole 
spin parameter $a_\ast$. We used as lower limit for the bolometric luminosity  
$\sim 7\times 10^{44}$ erg s$^{-1}$, 
as found by \citet{gal04} from the observed free-free emission, and the Eddington luminosity $L_{\rm Edd}$ as upper limit.

The Bardeen-Petterson radius was obtained from either equation (6) or (8), 
depending if it obeys the diffusive or wave regimes, respectively.
 
The calculations were performed 
for the disk models A  and B defined in Table 1, considering 
several values of the black hole spin and the extreme values of the 
accretion rate and $\alpha$ parameter, the latter depending also on 
the spin value. The results show that the Bardeen-Petterson radius 
for NGC\,1068 depends weakly on the accretion rate, being limited 
by the variation of the black hole spin and $\alpha$ parameters 
between 10$^{-5}$ and 10$^{-4}$ pc (about 20 and 200 $R\mathrm{g}$ 
respectively) as shown in left panel of Figure 2. 

For the lower limit of the $\alpha$ parameter and both black hole 
masses, $R_{\rm BP} = R_{\rm BP}^{\rm w}$ when $|a_\ast|> 0.1$. For 
the upper limit of $\alpha$, $R_{\rm BP} = R_{\rm BP}^{\rm diff}$ for   
$ a_\ast < 0.1$, turning gradually into $R_{\rm BP}^{\rm w}$ for 
larger values of $a_\ast$ at a rate that depends on the $\alpha$ 
parameter.

We calculated the alignment time-scale  as a function of the 
black hole spin by solving the integral given in equation (3), 
using as integration limits the Bardeen-Petterson radius and 
an arbitrary outer radius. The results, which are independent 
of the outer radius value, are presented in the right panel of 
the Figure 2. 

The alignment time-scale turned out to vary between 100 and 10$^5$ yr, 
while the lifetime of the radio jet 
in NGC\,1068 is $\la 1.5\times10^{5}$ yr \citep{cap99}, similar to the lifetimes of  AGN activity. 
Therefore, it indicates that the Bardeen-Petterson effect can 
perfectly warp the inner part of the accretion disk.

It is important to emphasize that warped accretion disks produced 
by the Bardeen-Petterson effect can be probed by features in 
relativistically broadened emission iron-line profiles \citep{fra05} 
but unfortunately, since NGC\,1068 is a type 2 Seyfert galaxy, 
relativistically broadened iron-lines are unlikely to be observed due 
to the obscuring dust torus.

\begin{figure*}
%%\epsscale{.80}
\plotone{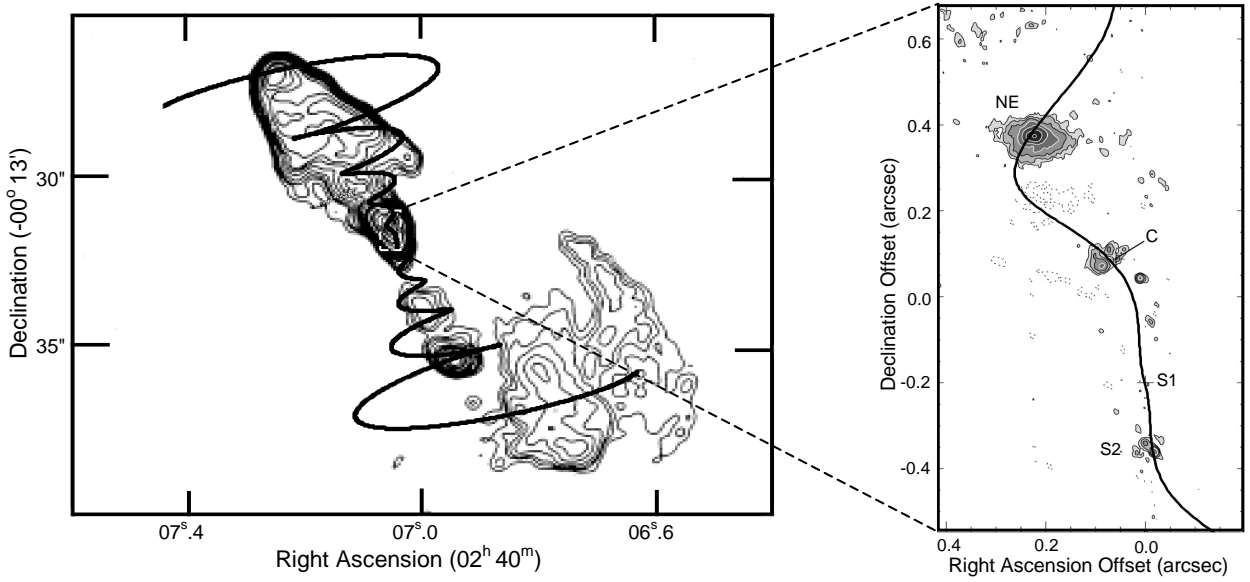}
\caption{Snapshot of the changes in the jet orientation due to the Bardeen-Petterson 
effect. {\it Left panel}: Kiloparsec scale jet of NGC\,1068 at 5 GHz \citep{wiul87}. 
Solid line indicates the model prediction from equations (9) and (10), assuming 
$\theta=80\degr$, $v_\mathrm{jet}=0.17c$, $t_0=-9800$ yr, 
$T_\mathrm{align}=7580$ yr, $\varphi_0=40\degr$ and $\eta=17\degr$. 
{\it Right panel}: Comparison between the map at 1.4 GHz \citep{gal04} with the 
same model shown in the left panel.}
\label{Jet}
\end{figure*}

\placefigure{Jet}

\section{Probing alignment from jet kinematics}

The current angular resolution provided by interferometric techniques 
is not capable to resolve structures with sizes comparable to the 
Bardeen-Petterson radius in NGC\,1068. However, the orientation of the 
inner part of the disk can be traced from the jet, which is usually 
thought (sometimes supported by observations; \citealt{ray96,jon00}) to be ejected in the perpendicular direction. 
In NGC\,1068, the jet is 
not continuous but presents discrete features, the upper limit to 
their expanding velocities being 0.17$c$ \citep{gal96c}.

The Bardeen-Petterson effect forces the disk to align gradually with 
the black hole, producing precession and a progressive change in the 
jet direction \citep{cap04}. According to \citet{scfe96}, the solution 
of equation (3) gives an exponential time variation for the inclination 
angle between black hole spin and the jet direction, as well as for the 
precession period $P_\mathrm{prec}$:

\begin{equation}
\varphi(t) = \varphi_0e^{-(t-t_0)/T_\mathrm{align}}
\end{equation}

\begin{equation}
P_\mathrm{prec}(t)=P_\mathrm{0}e^{-(t-t_0)/T_\mathrm{align}}
\end{equation}
\\where $\varphi_0$ and $P_\mathrm{0}$ are, respectively, 
the inclination angle and precession period at time $t_0$ 
when the disk was formed ($t_0\leq 0$, measured in the past 
from the present time). \citet{scfe96} found that the 
timescales for precession and realignment are identical, 
implying that $P_0=T_\mathrm{align}$, which will be also 
consider in our approach.

We assumed that the jet originated at time $t_0$, afterwards each plasma element was ejected with constant speed %%@
$v_\mathrm{jet}$ at a time $t$, in a direction that forms an angle $\varphi (t)$ with the black hole spin axis, in a plane %%@
that rotates with velocity  $\omega(t) = 2\pi/P_{\rm prec}(t)$.
We introduced two additional parameters: the viewing  and position angle
of the black hole spin in relation to the line of sight $\theta$ and on the plane of the sky $\eta$. 

In order to compare our model with the observations, we calculated 
the right ascension and declination offsets ($\Delta\alpha$ and 
$\Delta\delta$ respectively) of each jet element as a function of 
time through:

\begin{equation}  
\Delta\alpha(t) = \frac {v_\mathrm{jet}t}{D}\left[A(t)\cos\eta+B(t)\sin\eta\right],
\end{equation} 

\begin{equation}  
\Delta\delta(t) = \frac {v_\mathrm{jet}t}{D}\left[-A(t)\sin\eta+B(t)\cos\eta\right],
\end{equation} 
\\where $D$ is the distance to the observer, $A(t)=\sin\varphi (t)\cos\omega t$ and 
$B(t) = \sin\varphi (t)\sin\omega t\cos\theta+\cos\varphi (t)\sin\theta$. 

We found solutions for several combinations of the input 
parameters, in the range $25\degr\la\varphi_0\la45\degr$, $70\degr\la\theta\la90\degr$ 
and  $14\degr\la\eta\la20\degr$. The parameters $t_0$ and $T_\mathrm{align}$ are 
scaled by the jet velocity, ranging between 
$-1.7\times 10^5\la t_0$ (yr)$\la -9600$ and 
$7500\la T_\mathrm{align}$ (yr)$ \la 1.3\times 10^5$ for 
$0.01c\la v_{\rm jet}\la 0.17c$ \citep{wiul87,gal96c}. The limits found for 
$T_\mathrm{align}$ are perfectly compatible with those derived in last section.

In Figure 3 we present the comparison between the parsec and kiloparsec 
scale radio maps \citep{wiul87,gal04} and a model with parameters 
$\theta=80\degr$, $v_\mathrm{jet}=0.17c$, $t_0=-9800$ yr, 
$T_\mathrm{align}=7580$ yr, $\varphi_0=40\degr$ and 
$\eta=17\degr$. We can see that our simple kinematic approach 
reproduces satisfactorily the inverted S-shape of the kiloparsec jet, 
as well as the location of its jet components; at parsec-scales, 
the position of the jet components C and NE, as well as the 
counterjet knot S2 are also well-reproduced by the same 
model. As it can be seen in the left panel of Figure 3, 
at larger distances from the core, the full range of position 
angles provided by our precessing model is not found in the observational 
data, in the sense that the amplitude of the helix towards 
the northern and southern regions are systematically larger 
than the observed jet aperture. This may be due to the implicit 
assumption of accretion disk rigid-body precession during all the 
jet-time evolution, which might not be totally true, specially 
close to the initial time $t_0$. In fact, numerical simulations 
have shown a period of differential precession preceding the 
rigid-body configuration \citep{nepa00,fran05}. In addition, 
the galactic medium will influence the jet propagation, as can 
be seen from the [\ion{O}{3}] and mid-infrared images of 
NGC\,1068 \citep{gal96a,cap97,gal03}, showing the existence of 
jet-ambient interaction.

In all cases considered in this work, the inner disk is 
not completely aligned with the equator of the black hole since 
$\varphi\sim 11\degr$ at the present time. Interestingly, 
\citet{gal03} proposed the existence of a slight tilt of $\sim 15\degr$ 
between the Compton thick central absorber and the molecular disc 
in order to reproduce the observed distribution of H$_2$ and CO 
emission in this object.

\section{Conclusions}

We 
studied the possibility that the Bardeen-Petterson effect is responsible 
for the warping of the accretion disk in NGC\,1068, as  derived from the radio continuum 
and water maser observations (e.g., \citealt{grgw97,gal04}). Such mechanism arises from the frame dragging 
produced by a Kerr black hole whose rotation axis is not parallel to 
that of the accretion disk.

Using an analytical approach, similar to that suggested by \citet{scfe96} 
and \citet{naar99}, we calculated the Bardeen-Petterson radius and the 
alignment time-scale between the accretion disk and the equator 
plane of the black hole for different values of the black hole spin. We 
found that the Bardeen-Petterson radius is roughly limited between 
10$^{-5}$ and 10$^{-4}$ pc, while the alignment time-scale ranges 
from about 100 to $10^5$ yr. Those estimates are perfectly 
compatible with the upper limit to the AGN lifetime, derived from the 
jet kinematics.

We also showed that the general form of the parsec and kiloparsec scale 
jets can be reproduced by the model in which the jet is precessing with 
a period $P_{\rm prec}$ and aperture angle $\varphi$ that decrease 
exponentially with a time-scale $T_{\rm align}$, as expected from the 
Bardeen-Petterson effect.


%% Included in this acknowledgments section are examples of the
%% AASTeX hypertext markup commands. Use \url without the optional [HREF]
%% argument when you want to print the url directly in the text. Otherwise,
%% use either \url or \anchor, with the HREF as the first argument and the
%% text to be printed in the second.

\acknowledgments

This work was supported by the Brazilian Agencies FAPESP and CNPq. We 
thank the anonymous referee for her/his useful comments.

\clearpage

\clearpage

\begin{deluxetable}{lcc}
\tabletypesize{\scriptsize}
%%\rotate
\tablecaption{Physical parameters of the disk.\label{tbl-1}}
\tablewidth{0pt}
\tablehead{
\colhead{Parameter} & \colhead{Model A} & \colhead{Model B}
}
\startdata
$M_\mathrm{BH}$ (M$_\sun$)       & (1.2~$\pm$~0.1)$\times 10^7$      & (8.0~$\pm$~0.3)$\times 10^6$    \\
$s$                              & -1.05~$\pm$~0.10                  & -1.05~$\pm$~0.10                \\
$M_\mathrm{d}$(r=1pc) (M$_\sun$) & (9.4~$\pm$~1.6)$\times 10^6$      & (8.6~$\pm$~0.6)$\times 10^6$    \\
$\Sigma_0$ (g cm$^{-2}$)         & (1.06~$\pm$~0.23)$\times 10^9$    & (1.49~$\pm$~0.18)$\times 10^9$  \\
$L_\mathrm{Edd}$ (erg s$^{-1}$)  & (1.5~$\pm$~0.1)$\times 10^{45}$   & (1.0~$\pm$~0.3)$\times 10^{45}$ \\
$\dot{M}/\dot{M}_\mathrm{Edd}$   & 0.46 - 1.0                        & 0.71 - 1.0                        \\
$\alpha\epsilon(a_\ast)$         & (4.4 - 9.4)$\times 10^{-4}$       & (4.4 - 6.3)$\times 10^{-4}$       \\
\enddata

\tablecomments{The quoted errors in $M_\mathrm{BH}$ and $s$ are 
given by \citet{hure02} and \citet{lobe03}, while for $\Sigma_0$ 
were obtained from error propagation.}

\end{deluxetable}

\end{document}